\documentclass[12pt,preprint]{aastex}
\usepackage{graphicx}
\usepackage{epsfig}
\usepackage{amsmath,amssymb}
\usepackage{lscape}
\usepackage{verbatim}
\usepackage{rotating}
\usepackage{subfigure}
\usepackage{epstopdf}

\def\lapprox{\lower.4ex\hbox{$\;\buildrel <\over{\scriptstyle\sim}\;$}}
\def\gapprox{\lower.4ex\hbox{$\;\buildrel >\over{\scriptstyle\sim}\;$}}

\shorttitle{Hot Halo Around NGC 1961}
\shortauthors{Anderson et al.}

\begin{document}
\title{Detection of a Hot Gaseous Halo Around the Giant Spiral Galaxy NGC 1961}
\author{Michael E. Anderson\altaffilmark{1}, Joel N. Bregman\altaffilmark{1}\\}
\altaffiltext{1}{Department of Astronomy, University of Michigan, Ann Arbor, MI 48109; 
michevan@umich.edu, jbregman@umich.edu}

\begin{abstract}
Hot gaseous halos are predicted around all large galaxies and are critically important for our understanding of galaxy formation, but they have never been detected at distances beyond a few kpc around a spiral galaxy. We used the Chandra ACIS-I instrument to search for diffuse X-ray emission around an ideal candidate galaxy: the isolated giant spiral NGC 1961. We observed four quadrants around the galaxy for 30 ks each, carefully subtracting background and point source emission, and found diffuse emission that appears to extend to 40-50 kpc. We fit $\beta$-models to the emission, and estimate a hot halo mass within 50 kpc of $5\times10^9 M_{\odot}$. When this profile is extrapolated to 500 kpc (the approximate virial radius), the implied hot halo mass is $1-3\times10^{11} M_{\odot}$. These mass estimates assume a gas metallicity of $Z = 0.5 Z_{\odot}$. This galaxy's hot halo is a large reservoir of gas, but falls significantly below observational upper limits set by pervious searches, and suggests that NGC 1961 is missing 75\% of its baryons relative to the cosmic mean, which would tentatively place it below an extrapolation of the baryon Tully-Fisher relationship of less massive galaxies. The cooling rate of the gas is no more than 0.4 $M_{\odot}$/year, more than an order of magnitude below the gas consumption rate through star formation.  We discuss the implications of this halo for galaxy formation models.
\end{abstract}


\maketitle

\section{Introduction}

Hot gaseous halos around galaxies have been an important prediction of galaxy formation models since White and Rees (1978). Theory predicts these hot halos form as matter accretes onto the dark matter halo and the baryons shock to the virial temperature (White and Frenk 1991, also see the review by Benson 2010). Depending on the details of the assumed pre-heating, heating from galactic feedback, and cooling rates, these hot halos are often predicted to contain as much or more baryonic mass as the galaxies within the halos (Sommer-Larsen 2006, Fukugita and Peebles 2006), making them cosmologically important as reservoirs of the ``missing baryons'' from galaxies (although also see Anderson and Bregman 2010). The hot halo is also thought to produce the galactic color-magnitude bimodality (Dekel and Birnboim 2006) and to help explain galactic ``down-sizing'' in the star formation history (Bower et al. 2006, De Lucia et al. 2006). 

Hot halos have been extensively observed in soft X-rays (roughly 0.5 - 2 keV) around early-type galaxies (Forman, Jones, and Tucker 1985, O'Sullivan, Forbes, and Ponman 2001, Mulchaey and Jeltema 2010). The halos are typically luminous ($L_{\text{X},0.5-2 \text{ keV}} \sim 10^{39} - 10^{41}$ for non-BCG ellipticals), mass-dependent (for most definitions of $L_{\text{X}}$ and $L_{\text{K}}$, $L_{\text{X}} \propto L_{\text{K}}^2$) and are often visible out to many tens of kpc. But these halos are difficult to connect to the formation of the galaxies because coronal gas can also be produced in the mergers and associated star formation that occurred when the galaxy became elliptical (Read and Ponman 1998), and because it is difficult to disentangle halo gas with the intergroup medium (IGM) in which most large ellipticals reside (Dressler 1980).

In contrast, hot halos around quiescent disk galaxies should be much more direct tracers of the galaxy formation process. While the morphology-density relation makes it difficult to disentangle elliptical galaxies from their dense environments, it also ensures a large supply of isolated spiral galaxies in low-density environments. Late-type disks are destroyed by strong mergers (e.g. Robertson et al. 2006), and it is easy to identify and exclude starbursting galaxies, so it should be straightforward to search for hot halos around quiescent isolated spirals, and to connect these halos to models of galaxy formation.

Unfortunately the search for extended soft X-ray emission around isolated spirals has so far been unsuccessful. There are several detections of emission a few kpc above the disk (Strickland et al. 2004a, Li et al. 2006, T\"{u}llmann et al. 2006, Rasmussen et al. 2009, Owen and Warwick 2009, Yamasaki et al. 2009), but these observations are linked to the star formation in the galaxy and probably represent galactic fountains. In terms of more extended emission, Li, Wang, and Hameed (2007) observe gas around the Sombrero galaxy out to 20 kpc, but this galaxy is significantly bulge-dominated, and the extended emission has been linked to a galactic bulge-driven wind. Finally, Pedersen et al. (2006) claimed to detect extended hot halo emission around NGC 5746, but this emission disappeared after subsequent re-analysis with newer calibration files (Rasmussen et al. 2009). 

A recent paper (Crain et al. 2010b) attributes these detections of extended emission to galactic coronae, instead of the standard explanation of the emission as a fountain or a wind originating from within the galaxy. This interpretation is in disagreement with the standard understanding of galactic fountains in spiral galaxies, but regardless of interpretation it still is true that no hot halo has been detected around a disk galaxy at a radius of more than a few kpc. 

In this paper, we present an analysis of \textit{Chandra} ACIS-I observations of the environs of the extremely massive spiral galaxy NGC 1961, in which we detect X-ray emission out to at least 40 kpc and attribute the emission to a hot halo. The outline of the paper is as follows. In $\S$2, we discuss the properties of NGC 1961 and the details of our observation. In $\S$3, we discuss the reduction of the data and explain various approaches to flat-fielding we adopted in our analysis. In $\S$4 we present the spatial analysis, and in $\S$5 we present the spectral results. Finally, in $\S$6 we discuss the derived properties of the hot halo and in $\S7$ we place them in the context of galaxy formation.

\section{Observation}
 This galaxy is one of the most massive spiral galaxies known (Rubin et al. 1979), with a maximum (deprojected) HI circular velocity of $402$ km s$^{-1}$ at 34 kpc (Haan et al. 2008). These authors fit for an inclination angle of $42.6^{\circ}$, close to the LEDA value of $47^{\circ}$. A recent paper (Combes et al. 2009) makes a plausibility argument for $i \approx 65^{\circ}$, which would reduce the circular velocity by 25\%. We use the more standard inclination in this paper, but if the higher inclination is correct, the total mass of the galaxy would be $0.75^{-3} = 2.4$ times smaller. While this is  a significant difference, it does not change our final conclusion.
 
If the relation $M_{\text{dyn}} \propto V_{\text{max}}^3$ holds at these velocities, we can compare NGC 1961 to the Milky Way ($v_{\text{circ}} = 220$ km s$^{-1}$) and infer that NGC 1961 has six times the dynamical mass of the Milky Way. Similarly, its 2MASS K-band magnitude is -26.0, which for an assumed mass-to-light ratio of 0.6 (Bell and de Jong 2001) corresponds to a stellar mass of $3\times10^{11} M_{\odot}$ (which is also six times the stellar mass of the Milky Way). Extrapolating from the $L_X - L_K$ relation for elliptical galaxies, we therefore expect NGC 1961 to have an unusually bright X-ray halo ($L_{0.5-2 \text{ keV}} \approx 1\times10^{41}$ erg s$^{-1}$ for the diffuse emission), making this galaxy an ideal target for identifying extended X-ray emission.  
 
The virial radius of the Milky Way is $\sim 250$ kpc (Shattow and Loeb 2009, Klypin, Zhao, and Somerville 2002), so by extension the virial radius of NGC 1961 would be around 450 kpc\footnote{We use the form for the virial radius defined in terms of the critical density $(\rho_{\text{vir}} \approx 200 \rho_c)$.}. Within this radius, NGC 1961 has several, much smaller, companions (Gottesman et al. 2002, Haan et al. 2008), including three dwarfs ($M_{\text{HI}} < 10^9 M_{\odot}$) at 120, 140, and 160 kpc, and several slightly larger galaxies at 200-500 kpc distances. It is therefore the dominant galaxy in a small group, but no IGM emission is observed. We adopt for this galaxy a distance of 56 Mpc (NASA Extragalactic Database average), which matches independent measurements of distances to other galaxies in the group (Gottesman et al. 2002), and is probably uncertain to 10\%.  At this distance, 1 arcminute corresponds to 16 kpc. 

Our observing strategy was to use a 2$\times$2 mosaic with the ACIS-I array on the Chandra X-ray Observatory, which allowed us to sample the extended emission out to about 260 kpc (17') - roughly 2/3 of the virial radius (see Figure 1). The observations (obs ids 10528-10531) were approved for 35 ks each, and ranged from 31.75-33.25 ks of good time. We also observed two background fields (obs ids 10532,10533) for 10.14 and 10.02 ks. All observations were taken in VFAINT mode with ACIS-I.

The data were processed using CIAO, version 4.1.2, and the latest calibration files. Data taken during flares were excised, but the observations were remarkably clean, with only about 0.4 ks of bad time to remove from the total integration time for each observation. A 0.6-6 keV image was produced for point source detection (using the WAVDETECT algorithm in CIAO). For the rest of the analysis, we use a 0.6-2.0 keV image for observations of the hot halo, and a 2.0-6.0 keV image for constraints on contamination from emission from X-ray binaries, as discussed in section 3.4.

\section{Flat-fielding}

This diffuse X-ray emission is very faint, even from such a large galaxy, which is why it has posed such an observational challenge. Therefore the most difficult and important part of the data reduction is the flat-fielding procedure. We tried several different methods of flat-fielding and background subtraction, described below. We eventually developed our own method of flat-fielding the observations (section 3.3), which should be generalizable to future observations of faint diffuse emission around nearby galaxies (section 7.1).

\subsection{Using Background Frames}

Our original intent was to use images taken a degree off-axis from the galaxy as background images which could be subtracted pixel-by-pixel from the halo images, accomplishing both flat-fielding and background subtraction at once. We took two background pointings for this purpose. The former (obs id 10532) was pointed at blank sky 17.3 degrees away from the galaxy for 10.14 ks, and was taken 29 days before the science pointings. The latter (obs id 10533) was pointed at blank sky 0.9 degrees away from the galaxy for 10.02 ks, 7.5 days before the science pointings began. 

Unfortunately these background images proved unsuitable for background subtraction because the count rate across the image was different than the count rates in the source images. After processing and point source subtraction, the mean 0.6-2 keV count rate in the background images is 0.31 count s$^{-1}$ compared to 0.36 count s$^{-1}$ in the source images. This 17\% discrepancy is comparable to the total signal from hot halo emission integrated out to 500 kpc (see section 6), and probably stems from variations either in the unresolved X-ray background of the images or in variations in the solar X-ray flux between the background observations and the source observations. In either case, the discrepancy was large enough that we were unable to use the background images for background subtraction. Similarly, the standard background images are even more spatially and temporally separate from our source images and could not be used either.

\subsection{In-field Subtraction, Modeling the Background}
We therefore decided to use in-field background subtraction instead. In-field subtraction is not ideal for our project, since we are trying to measure diffuse emission that, in principle, could fill the entire field of view. However, the diffuse emission is only detectable out to a few arcminutes from the galaxy, and that even over most of that range the background is larger than the signal. Thus it is possible to compute the background in-field, and use this for background subtraction.

We first attempted to model the background. We assumed the 0.6 - 2.0 keV background consisted of two components: a vignetted component from the diffuse X-ray background (which we assumed would be proportional to the effective throughput at each pixel, computed using the exposure map), and an unvignetted component from particles in the Solar wind. In each observation, we sampled the background at various points across the detector (away from the direction of the galaxy), and fit the background to a linear combination of these two components. 

This procedure worked adequately for two of the four observations (10528 and 10530), but the other two had large-scale variations in the background across the image or other un-modeled effects and therefore did not yield reasonable fits to the two background components.

\subsection{In-field Subtraction, Conjugate Technique}

We found more success with in-field conjugate subtraction. The idea is to assume the vignetted background is azimuthally symmetric over large scales around the aimpoint of the image (located on the ACIS-I3 chip at approximately (x,y) = (974,969) in detector coordinates\footnote{Chandra Proposers' Observatory Guide, version 13.0, http://cxc.harvard.edu/proposer/POG/html/index.html}). We selected ``conjugate'' background regions in the image of the same shape and size as the source, but at the opposite position angle from the aimpoint of the image. We then subtracted each conjugate region from the corresponding source region (see Figure 2). 

As shown in Figure 2, we faced different geometrical combinations of source, conjugate, and aimpoint for each observation. For observation 10528, the galaxy is just off the edge of the boundary between the I0 and I1 chips, the aimpoint is on the opposite side of the center of the ACIS-I array, and so the conjugate point is 170 arcseconds beyond the edge of the I3/I4 chips. We therefore had to choose annuli at the same distance from the aimpoint, but at angles of $120^{\circ}$ and $240^{\circ}$ instead of $180^{\circ}$ to find conjugate regions on the detector for the source photons from the inner 170 arcseconds (45 kpc) around the galaxy in this observation. For observation 10529, the galaxy is just off the edge of the I0/I2 chips, the aimpoint is again on the opposite side of the center of the array, and the conjugate point is 130 arcseconds beyond the edge of the I3/I1 chips. We therefore had to choose annuli at the same distance from the aimpoint, but at angles of $120^{\circ}$ and $240^{\circ}$ instead of $180^{\circ}$ to find conjugate regions on the detector for the source photons from the inner 130 arcseconds (35 kpc) around the galaxy in this observation.  For observation 10530, the galaxy is 200 arcsec off the edge of the I1/I3 chips, but the aimpoint is now on the same side of the center of the array, so the conjugate point is close to the edge of the I0/I2 chips. But since the galaxy is off the detector, we do not have a measurement of the source for the inner 200 arcseconds (50 kpc) around the galaxy in this observation. For observation 10531, the galaxy is on the I3 chip, near the I2 chip, 270 arcseconds from the aimpoint. The conjugate point is also on the detector, on the boundary between the I0/I1 chips. So for this observation, we have measurements of the source and background emission out to 270 arcseconds (75 kpc); beyond this radius the background and source annuli begin to overlap. We excluded the data beyond this radius for observation 10531 for the rest of the analysis, but did also repeat the analysis with these points included, and found that they have no effect on the results since the halo emission has disappeared by 50 or 60 kpc. 

This yielded similar backgrounds to the results of the modeling for 10528 and 10530, but the results were much better for 10529 and 10531, so we adopted this approach for the rest of the analysis. To help verify the reliability of this technique, we also tested the conjugate technique 90$^{\circ}$ on either side of the source, and obtained zero signal.

\subsection{Point Sources}

It is critically important to be sure we are measuring the hot diffuse emission and not a collection of X-ray binaries in and around the galaxy, whose surface density also falls off with radius like the halo gas. The first step to ensuring a clean measurement is the automated point source removal using WAVDETECT, described above, which removed six point sources within the inner 50 arcseconds, with the faintest point source having a luminosity of $ L_{\text{0.6-2 keV}} \sim 3\times10^{38}$ erg s$^{-1}$ if at the assumed distance of 56 Mpc. One of these point sources falls on the galactic nucleus, which hosts a low-luminosity AGN (Roberts and Warwick 2000) for which we measured a luminosity of $1\times10^{40}$ erg s$^{-1}$. 

To estimate the contribution of point source emission to the surface brightness profile, we extracted and reduced an image of the 2-6 keV emission, using the identical procedure as we used for the 0.6-2 keV images. We expect no contribution from the $< 1$ keV gas in this higher-energy band, so all the emission should come from point sources in the galaxy or the background. Using the in-field conjugate subtraction technique, we subtracted the background emission and derived radial surface brightness profiles for the 2-6 keV galactic emission. We attributed all this emission to unresolved point sources. 

Irwin, Athey, and Bregman (2003) found that the integrated emission from low-mass X-ray binaries (which dominate the point source emission over most of the region in our analysis) has a universal spectrum that can be fit with a power-law distribution with a slope $\Gamma = 1.56$. Using this slope and accounting for absorption, we find that each unit of 2-6 keV emission corresponds to 1.7 units of 0.6-2 keV emission. We scaled the 2-6 keV galactic emission by a factor of 1.7 and subtracted this from the 0.6-2 keV emission to remove the point source contribution. We stopped the unresolved point source removal after reaching the average background 2-6 keV surface brightness for each observation; this occurred at 32 kpc for observation 10528, 30 kpc for observation 10529, and 25 kpc for observation 10531 (observation 10530 has the galaxy 50 kpc off the edge of the detector, so there is no contribution from galactic point sources. The total amount of emission due to unresolved point sources was $\approx 1.5 \times 10^{40}$ erg s$^{-1}$, about 10-20\% of the total emission. 

We attempted to verify this result by assuming the six point sources detected with WAVDETECT represent a complete sample down to $3\times10^{38}$ erg s$^{-1}$. Using the {\it Chandra} point source number counts of Kim et al. (2007), we expect 0.6 background point sources above our flux limit in the (50 arcsecond radius) aperture, so one of these six sources is likely actually an unrelated background object (and one is the central low-luminosity AGN). We applied the luminosity function of Grimm, Gilfanov, and Sunyaev (2002), which was calibrated using Galactic LMXBs and HMXBs, to the five non-nuclear point sources. We adopted a power-law function for $N(>L)$ with a slope of $\alpha = -0.3$. The ratio of the total luminosity (with a lower-luminosity cut-off at $10^{36}$ erg s$^{-1}$ to the luminosity above $3\times10^{38}$ erg s$^{-1}$ is 5.0, and the total 0.6-2 keV luminosity of our five point sources is $7.0 \times10^{39}$ erg s$^{-1}$, so we expect a point source luminosity of $3.5\times10^{40}$ erg s$^{-1}$. This figure is about twice as high as the point source luminosity we infer from the 2-6 keV flux, although the likely presence of a background point source is probably causing us to overestimate the background using this method.  

If we assume that all five detected non-nuclear point sources within 50 arcseconds (13 kpc) are associated with the galaxy, then our method of scaling from the high-energy emission corresponds to a flat power-law slope of $\alpha = -0.15$ for $N(>L)$. This slope is $1.4\sigma$ away from the best-fit slope to the LMXB luminosity function in Grimm, Gilfanov, and Sunyaev (2002). It does seem more likely for NGC 1961 to have a different point source luminosity function than the Milky Way instead of a different X-ray spectrum for its point sources. Supporting this result, a flattening of the point source luminosity function below a few $\times 10^{37}$ erg s$^{-1}$ is also observed in Centaurus A (Voss et al. 2009). 

We also included a correction for X-ray emission from stars at a level fainter then $10^{36}$ erg s$^{-1}$. This emission seems to scale with total stellar mass (inferred from the K-band luminosity), at least for old stellar populations (Revnivtsev et al. 2008). While NGC 1961 is a late-type spiral and therefore not likely to be dominated by an old stellar population, the Revnivtsev et al. scaling is the best available at present for accounting for this emission. We took the K-band radial surface brightness profile for NGC 1961 from 2MASS (Jarrett et al. 2003), re-binned it to match our X-ray annuli, and assumed a K-band mass-to-light ratio of 0.6 to convert into stellar mass (Bell and de Jong 2001). We then applied the Revnivtsev et al. conversion between stellar mass and 0.5-2.0 keV X-ray luminosity to estimate the stellar X-ray emission. Finally, we multiplied the predicted X-ray emission by 73\% to account for absorption by the Galactic Hydrogen column in the direction of NGC 1961. We find that the stellar emission is never a significant fraction of the total X-ray emission, even at very small radii; the total 0.5-2.0 keV emission from stars is $2.6\times10^{39}$ erg s$^{-1}$, or about 5\% of the total luminosity in the central region. The K-band half-light radius is 35 arcsec (9.6 kpc), and the K-band surface brightness reaches the 2MASS $1\sigma$ background at about 78 arcsec (21 kpc) -- a much more concentrated profile than the X-ray emission.  

\section{Radial Surface Brightness Profile}

Our primary interest is in deriving a radial surface brightness profile for the hot gas around NGC 1961. The emission is very faint, so we have to make a few critical assumptions in  order to make any progress in parameterizing the surface brightness profile. Most importantly, we expect the halo to be roughly spherical, so we compute the surface brightness profile in circular annuli around the galaxy. Starting with just this assumption, in Figure 3 we present background-subtracted radial surface brightness profiles for our four images of the halo around NGC 1961. We chose annuli such that each source annulus contains at least 20 photons, although we also tried annuli of constant radius and found no significant differences in the shape of the profile. 

Inspecting Figure 3, emission seems visible out to about $160$ arcsec, corresponding to about $40$ kpc. This is the key result of our paper, and it does not depend on any further assumptions or statistical techniques (and we find this excess in multiple observations/quadrants). Some of the interior emission is due to X-ray binaries and a low-luminosity AGN, but we attribute the bulk of the emission at 40 kpc to diffuse gas around the galaxy. We discuss in section 7.3 the possibility that this emission results from some internal galactic process (i.e. a supernova-driven wind or galactic fountain), but we conclude this possibility is unlikely. 

In the rest of this section, we will attempt to quantify the level of emission above the background out to 40 kpc, by introducing parametric fits to the surface brightness profile and then by smoothing the background.

\subsection{Parametric Fitting with the $\beta$-model}

Our assumed parametric form for the surface brightness profile is the class of models knows as the $\beta$-models. These models parameterize the surface brightness as

\[S(r) = S_0 \left[ 1 +\left(\frac{r}{r_0}\right)^2 \right]^{0.5-3\beta}\]

which, if the gas is isothermal and has a constant metallicity, corresponds to a density distribution of the form

\[n(r) = n_0 \left[1 + \left(\frac{r}{r_0}\right)^2\right]^{-1.5\beta}\]

These models provide good fits to hot gas around elliptical galaxies (Forman, Jones, and Tucker 1985) as well as the hot gas in galaxy groups and clusters (Sarazin 1986). We quantified fits to the $\beta$-models using $\chi^2$-minimization and we binned the data to have at least 20 photons per radial bin. Using the radial surface brightness profiles, we attempt to exclude specific choices of $\beta$-model, and the best-fit models are the profiles which can be excluded at the lowest confidence. We want to be able to exclude at less than 95\% confidence for the model to be considered statistically acceptable. 

We fit the surface brightness profiles to the $\beta$-profiles and solved for $S_0$, $r_0$, and $\beta$ using $\chi^2$ minimization. We include annuli extending out to 370 arcsec in our fit, since this radius appears to enclose all the excess emission visible in the surface brightness profile (see Figure 3). However, varying this radius does not affect the result much. We also require a core radius of at least 1 kpc. It would be better if we did not have to constrain the core radius at all, but we have no observations within the core radius so an observational constraint is difficult. A 1 kpc core radius is very small for a hot gaseous halo around a galaxy of this size, so our constraint is at least still somewhat conservative.

We fit all four profiles simultaneously and to find a single set of parameters that worked for all four observations. The best-fit parameters were $S_0 =9.77\times10^{-8}$ count s$^{-1}$ cm$^{-2}$ arcsec$^{-2}$, $r_0 = 1.00$ kpc, $\beta = 0.47$, and the full range of acceptable fits is narrowly clustered around these values (see Figure 3). 

While we had at least 20 source photons in each radial bin, in the inner annuli where the galaxy is brighter than the background there are fewer than 20 background photons per bin. Additionally, from the size of the radial background variations at large radii, we can infer there are small-scale spatial variations in the background (probably due to unresolved point sources). We attempted to correct for these two effects by fitting a second-order polynomial to the surface brightness profile of the background. Quadratics were the lowest-order polynomials with acceptable fits to the background surface brightness profile (reduced $\chi^2 = 1.42$, $1.70$, $1.08$, and $1.02$, respectively, for 33, 33, 26, and 47 degrees of freedom). Observation 10529 has the least well-behaved background at small radii, and this is partially responsible for the lower values at small radii for this observation.

As noted above, our main conclusion -- that there is extended coronal emission around NGC 1961 out to at least 40 kpc -- does not depend on this smoothing techniques, and we can still get an acceptable fit to the data without any smoothing. Smoothing the background allows us to remove a principal source of error in our analysis, however, yielding a wider and more reliable range of acceptable fits. The data with smoothed background, as well as the acceptable fits, are shown in Figure 4. We take the fits to the smoothed data (unacceptable at less than 95\% confidence) as the fiducial range for the rest of the paper. Note that this range encompasses the entire range of acceptable fits to the unsmoothed data.

We present the smoothed surface brightness profile in log-log space in Figure 5. In this figure, we have subtracted out the estimated contribution from X-ray binaries and from stars, as discussed in section 3.4. For ease of visualization in log-log space, for this figure we have not subtracted out the smoothed background; rather, we indicate the level of the smoothed background for comparison. Again, we clearly detect  emission above the background out to 40-50 kpc, in multiple quadrants, and this emission is more extended than the emission from stars and X-ray binaries.

The parameters for the joint fit with the highest enclosed mass are ($S_0 = 3.85\times10^{-8}$ count s$^{-1}$ cm$^{-2}$ arcsec$^{-2}$, $\beta = 0.41$, $r_0 = 1.00$ kpc), and the parameters for the fit with the lowest enclosed mass are  ($S_0 =1.38\times10^{-8}$ count s$^{-1}$ cm$^{-2}$ arcsec$^{-2}$, $\beta = 0.54$, $r_0 = 4.04$ kpc).

\section{Spectral Fitting}
We also examined the spectrum of the source photons, to verify that the emission is consistent with strong metal emission lines atop a thermal bremsstrahlung continuum as expected (and not, for example, unresolved point sources), and to measure the temperature of the hot emitting gas. 

We examined the spectrum of observation 10531. We made a 0.5-6 keV image and removed point sources using WAVDETECT, then selected the inner 40 kpc (160 arcsec) a source aperture and a conjugate circle of the same size as a background region. This is the radius at which the signal approximately matches the background, and therefore should yield the spectrum with the optimal signal/noise ratio. 

Using XSPEC version 12.6.0, we fit various models to the data. The hot halo is expected to match an APEC model, and the X-ray binaries are expected to follow a $\Gamma = 1.56$ powerlaw, and both components should experience Galactic photoabsorption, so we fit the signal with an (APEC + powerlaw) $\times$ PHABS model. To account for the background (Galactic and extragalactic) and instrumental features, we followed Humphrey et al. (2011) and used an (absorbed) $\Gamma = 1.41$ powerlaw for the extragalactic background and two (unabsorbed) APEC models with Solar abundance and $kT = 0.07$ keV and $kT = 0.20$ keV for the Galactic background. We also included two zero-width Gaussian features at 1.77 and 2.2 keV to account for Si and Au emission lines from the instrument itself. The normalizations of all these components were allowed to float, but we fixed the hot halo metallicity at $Z = 0.5 Z_{\odot}$ and the Galactic column at $8.28\times10^{20}$ cm$^{-2}$ (Dickey and Lockman 1990). The best-fit model has a temperature for the hot halo component of $kT = 0.60^{+0.10}_{-0.09}$ keV and a normalization of $1.07^{+0.36}_{-0.27} \times 10^{-4}$. The $\Gamma = 1.56$ powerlaw component (corresponding to XRB emission from NGC 1961) had a normalization of $1.71^{+0.79}_{-1.71} \times 10^{-5}$. The $\chi^2$ was 119.5 for 122 degrees of freedom, corresponding to a reduced $\chi^2$ of 0.98. 

We tried a fit with an additional PHABS component to account for absorption within the disk of NGC 1961, but this component did not improve the fit at all. Letting the photoabsorbing column density float instead of fixing it at $8.28\times10^{20}$, we are not able to constrain meaningfully the value. Similarly, we also tried a fit with the hot halo metallicity left floating (instead of fixed at $0.5 Z_{\odot}$), and found the same halo temperature but no constraint on the metallicity. We were able to estimate the neutral Hydrogen column in NGC 1961 using the 21-cm observations of Haan et al. (2008). They find a clumpy distribution of column density ranging between $2-4 \times 10^{20}$ throughout the inner 20 kpc, with a sharp dropoff at larger radii. We chose $3\times10^{20}$ as a rough average column for the inner 20 kpc. A fit with this additional photoabsorption component on the APEC and the powerlaw models is equally acceptable and yields essentially the same temperature and normalization. Based on the Haan et al. (2008) column densities, however, we do apply a $\sim 10\%$ upwards correction to the X-ray surface brightness in Figures 3 and 4 and in the rest of the spatial analysis for bins within 20 kpc to account for the additional photoabsorption, although this has almost no effect on the final profile. 

In our aperture, we have approximately $767\pm65$ source photons out of a total of 2510 photons in the 0.5-6 keV energy range. Overall,  the results of the spectral fitting are in agreement with our theoretical expectations for the source of the emission. In particular, we expect a halo temperature  under 1 keV, and the APEC component to have a total emissivity about 5-10 times as large as the powerlaw component. Both these expectations are met, and additionally the APEC normalization from the spectral fit is consistent with the APEC normalization we infer from spatial fitting in the next section.

\section{Halo Mass}
We estimate the halo mass from the results of the spatial fit ($S_0$, $r_0$, and $\beta$) after removing individual and unresolved point sources. We integrate the $\beta$ model out to 50 kpc, after which radius the signal is down to below 20\% of the background. The enclosed source photon count rate is between 0.011-0.012 count s$^{-1}$. We then use the PIMMS utility (assuming an APEC emission model with $kT = 0.60$ keV and $Z = 0.5 Z_{\odot}$) to convert to an unabsorbed flux ($9.08-10.52 \times10^{-14}$ erg cm$^{-2}$ s$^{-1}$) and an unabsorbed luminosity (L$_{0.6-2} = 3.4-3.9\times10^{40}$ erg s$^{-1}$). We examined APEC models in XSPEC using the above parameters, and found that the acceptable fits all correspond to a normalization between $8-9\times10^{-5}$. This normalization can be converted into a central electron density ($n_0 = 1.2-2.8\times10^{-2}$ cm$^{-3}$) and an enclosed gas mass within 50 kpc of $4.9-5.2\times10^9 M_{\odot}$ after applying a 30\% upwards correction to account for the cosmological Helium fraction).

If we extrapolate this integration out to 500 kpc (the approximate virial radius of this galaxy), the count rate increases to 0.018-0.036 count s$^{-1}$, corresponding to an unabsorbed flux of  $1.50-3.06 \times10^{-13}$ erg cm$^{-2}$ s$^{-1}$) and an unabsorbed luminosity of L$_{0.6-2} = 5.6-11.5\times10^{40}$ erg s$^{-1}$. The enclosed gas mass within 500 kpc is $1.4-2.6\times10^{11} M_{\odot}$. The best-fit profile corresponds to an enclosed gas mass of $2.3\times10^{11} M_{\odot}$.

\subsection{Flattened Profiles}
We also examined joint and individual fits with two-component $\beta$-models. One component is the $\beta \sim 0.5$ profile described above, and another is a much flatter profile ($\beta \sim 0.35$) with much a larger core radius ($r_0 \sim 100$ kpc). This joint profile is motivated by theoretical work that suggests hot halos might have higher entropy than expected, which would flatten the profiles and allow them to contain more material (Crain et al. 2010a, Kaufmann et al. 2009, Guedes et al. 2011). In a previous paper (Anderson and Bregman 2010) we used a fiducial slope of $\beta = 0.35$ for flattened profiles, so we constrain the flattened profile to have that slope in our fit. We also fix the core radius at 50 kpc to reduce the parameter space, but our conclusion is not very sensitive to this parameter. We  choose logarithmically-spaced values for the normalization of the flattened component, and for each normalization, we search for values of $\beta$, $r_0$, and $S_0$ corresponding to the non-flattened ($\beta \sim 0.5$) profile, such that we minimize the total $\chi^2$ of the fit to the data of the sum of the flattened and non-flattened profiles. We fit the data out to radii of 500 arcseconds so as to incorporate constraints from beyond the region of visible emission.

The resulting minimum $\chi^2$ as a function of the normalization of the flattened profile component, and the strongest constraint (ruled out at 95\%) is a maximum normalization of $S_0 = 1.5\times10^{-10}$  count s$^{-1}$ cm$^{-2}$ arcsec$^{-2}$. This corresponds to a total count rate out to 500 kpc of $6.5\times10^{-2}$  counts s$^{-1}$, or an unabsorbed flux of $5.2 \times 10^{-13}$ erg s$^{-1}$ cm$^{-2}$, or a 0.6-2 keV luminosity of $2.0 \times 10^{41}$ erg s$^{-1}$. The normalization is $1.1 \times 10^{-3}$, corresponding to a hot halo mass in the flattened component of $7.4 \times 10^{11}$ $M_{\odot}$.

\section{Implications and Conclusion}

A full treatment of the implications of this measurement and this unusual galaxy is best reserved for future work, so here we comment briefly on a few salient points.

\subsection{Generalizing the Conjugate Subtraction Technique}

Given the historical observational difficulties in detecting diffuse emission at large radii around spiral galaxies, and our own experience and difficulties with this observation, we point out a few lessons learned that may be of use for future observers in this field. First, it is inadvisable to place the galaxy too close to the aimpoint of the telescope. There are numerous instrumental effects (vignetting, point source detectability, decreasing throughput, etc) that all vary with distance from the aimpoint; while corrections can be applied for each, it is very difficult to disentangle instrumental effects from diffuse emission at or below the surface brightness of the background. Second, it is also inadvisable to place the galaxy at the very edge of the detector (as we did for observations 10528 and 10529). While this configuration allows for conjugate subtraction to circumvent many of the radially varying instrumental effects, the throughput and vignetting effects are so large at the edge of the detector that it becomes difficult to see the galaxy. The optimal configuration is probably similar to our observation 10531, where the galaxy is placed a few arcminutes from the edge of the detector. This allows for conjugate subtraction while still retaining enough throughput to detect faint emission. The primary shortcoming of this configuration is that the conjugate annuli begin to overlap sooner than they would if the galaxy is at the edge of the detector. For observation 10531, we can only probe the inner 4.5 arcminutes of the halo, but the halo falls below the background at a radius of about 3 arcminutes, so we can still see all the useful emission. 

As described in section 4.1, we also found that a quadratic fit to the soft X-ray background as a function of radius is generally statistically acceptable. Using quadratic fits to the background for conjugate subtraction, our technique is able to detect emission down to about $20\%$ of the background for very faint and diffuse emission (see Figure 4).

\subsection{Halo Faintness and the Baryon Budget of NGC 1961}

As mentioned in section 1, it has been known that hot halos around spiral galaxies are underluminous in comparison to hot halos around ellipticals, but now that we have a detection of one, we can finally quantify this. Benson et al. (2000) examined a sample of three large spirals, of which the largest (NGC 4594, the Sombrero galaxy) is similar in mass but much more bulge-dominated than NGC 1961. They predict a bolometric hot halo luminosity within the inner 76 kpc of $6.9\pm0.5\times10^{41}$ erg s$^{-1}$ for a metal abundance of $0.3 Z_{\odot}$ (which would be 50\% higher for our assumed abundance of $0.5 Z_{\odot}$. This is not observed, but they do establish an observational upper limit of $4.4\pm2.8 \times10^{40}$ erg s$^{-1}$ for the emission at radii between 16 and 30 kpc. Our detection falls below this limit. We find an absorbed 0.6-2 keV luminosity (which is about half the bolometric luminosity for 0.6 keV gas) within the 16-30 kpc annulus of $5.2-6.5\times10^{39}$ erg s$^{-1}$, which is 75\% below their upper limit and only 2\% as luminous as the theoretical prediction.  The hypothetical flattened component would be even fainter within this annulus, with a luminosity of only $2.3\times10^{39}$ erg s$^{-1}$.

The baryon fraction of this galaxy is dominated by the stars: the 2MASS K$_S$-band total magnitude of 7.73 corresponds to an absolute K$_S$ magnitude of -26.0. Assuming a mass-to-light ratio of 0.6 (Bell and de Jong 2001), we infer a stellar mass of $3.1\times10^{11} M_{\odot}$. The cold gas component can be approximated by the HI mass, which is $4.7\times10^{10} M_{\odot}$ (Haan et al. 2008). We find a hot halo mass of $1.4-2.6\times10^{11} M_{\odot}$. So the baryon budget of this galaxy is approximately 6:1:3-5 stars : cold gas : hot gas (with the flattened profile the ratio is 6:1:14). Within 50 kpc, the baryon budget is approximately 60:10:1 stars : cold gas : hot gas. So, out to 50 kpc, most of the gas in NGC 1961 is cold, but if we are justified in integrating to the virial radius, the majority of the gas in the galaxy is in the hot phase, as predicted by theory for large galaxies. 

It is important to highlight the uncertainties in our measurement of the hot halo mass. Our $\beta$-model fitting has a 95\%-confidence uncertainty of a factor of three in the halo mass. A 10\% uncertainty in the distance to the galaxy would introduce a roughly 20\% additional uncertainty in the estimated mass, and our choice of 500 kpc for an outer radius of integration probably introduces another 10\% error relative to the true (unknown) virial radius. The assumptions of isothermality and azimuthal symmetry may not be true at larger radii as well. By far the biggest source of uncertainty, however, is the metallicity of the gas. At the virial temperature of this galaxy, X-ray emissivity is nearly linearly proportional to the gas metallicity, so if the gas is actually enriched to $1 Z_{\odot}$ instead of our assumed value of $0.5 Z_{\odot}$, the true halo mass is half the value we infer, and if the gas is actually $0.25 Z_{\odot}$, the true halo mass is twice the value we infer. A metallicity gradient in the hot halo would make an accurate mass estimate even more difficult. Obtaining a measurement of the metallicity of the hot halo gas around a spiral galaxy is critically important for accurately measuring the mass (as well as constraining the formation history of the halo).

For our value of $Z = 0.5 Z_{\odot}$, we can convert the baryon budget into a baryon fraction with an estimate of the total dark halo mass. The most recent measurement of the circular velocity finds an inclination-corrected HI velocity at 34 kpc of 402 km s$^{-1}$ (Haan et al. 2008), which is comparable to older measurements for this galaxy (e.g. Rubin et al. 1979). If this measurement is correct and the dark matter follows an NFW profile, the expected virial mass is $M_{200} = 2.3\times10^5 v_f^3 h^{-1} M_{\odot}$ (Navarro 1998). So NGC 1961 would have an inferred mass of $2.1\times10^{13} M_{\odot}$ - the mass of a medium-sized galaxy group. Even if the halo is not precisely NFW, the total mass seems unlikely to differ by more than 50\% or so. 

Since NGC 1961 is at the center of a poor group, the other galaxies in its group should also be added to the baryon budget. We added the HI masses for the six other candidate group members (Haan et al. 2008) and found a total HI mass of $7.5\times10^9 M_{\odot}$, which is less than a sixth of the HI mass of NGC 1961. We estimated the stellar mass of the six group members using their H-band absolute magnitudes from NED and a mass-to-light ratio of 0.6; this yields a total stellar mass for the group members of $3\times10^{10} M_{\odot}$, less than a tenth of the stellar mass of NGC 1961. NGC 1961 is therefore by far the dominant reservoir of baryons in its group, but we do include the small baryonic contributions from the other galaxies when we compute the baryon budget.

Within 500 kpc, the baryon fraction $f_b \equiv M_b / M_{\text{total}}$ is therefore 0.024-0.029 (or 0.051 with a flattened halo profile) - far less than the cosmological mean value $f_b = 0.171 \pm 0.006$ (Dunkley et al. 2009). This corresponds to a baryon fraction within $R_{500}$ of 0.023-0.033 for the single-component fit, and a maximum baryon fraction within $R_{500}$ of 0.043 for the two-component flattened $\beta$-model. Thus, within the virial radius NGC 1961 is missing over 75\% of its baryons, which is surprising since the missing baryon fraction is nearly always smaller in structures of this size (McGaugh et al. 2010). 

The baryonic Tully-Fisher relation (Stark, McGaugh, and Swaters 2009) predicts this galaxy should have a baryonic mass of $1.1\times10^{12} M_{\odot}$ (corresponding to a baryon fraction $f_b = 0.052$), so NGC 1961 is slightly below the BTF, but still potentially within the typical scatter about the relation. Including a flattened profile would definitely place this galaxy on the BTF, as would changing the gas metallicity to $0.2 Z_{\odot}$. Additionally, using the higher inclination angle of Combes et al. (2009) (discussed in section 2) would also bring NGC 1961 onto the BTF due to the steep dependence of this relation on $v_{\text{circ}}$.

\subsection{Halo Cooling Rates and Implications for Galaxy Formation}
We can estimate the cooling radius of this hot halo and the implied accretion rate onto the galaxy, which has implications for setting and regulating the star formation rate in the galaxy. We define the cooling radius as the radius for which the cooling time is 10 Gyr, using the expression for cooling time from Fukugita and Peebles (2006):

\begin{equation}
\tau(r) = \frac{1.5 nkT}{\Lambda n_e \left(n-n_e\right)} \approx \frac{1.5kT\times1.92}{\Lambda n_e \times 0.92}
\end{equation}

where the latter expression assumes primeval Helium abundance so the total particle density $n = 1.92 n_e$. For $T = 10^{6.85}$ K and $Z = 0.5 Z_{\odot}$, $\Lambda = 10^{-22.85}$ erg cm$^{3}$ s$^{-1}$ (Sutherland and Dopita 1993). Thus the cooling radius occurs at $n_e = 6.8\times10^{-4}$ cm$^{-3}$. For the range of best-fit $\beta$-model profiles listed above, this corresponds to a cooling radius between 17.8 and 18.2 kpc, and an interior hot halo mass of $8.9-10.2\times10^8 M_{\odot}$. It is  difficult to estimate the accretion rate onto the disk from this hot halo, since the heating rate is unconstrained, but we can make an order-of-magnitude estimate by dividing the hot gas thermal energy within the 10 Gyr cooling radius by the luminosity within that radius; this yields a cooling time of 2.0-2.4 Gyr for material within the cooling radius, or an effective cooling rate of $0.4 M_{\odot}$ year$^{-1}$. In contrast, we can estimate the star formation rate in NGC 1961 from the total H$\alpha$ luminosity ($7.6 \pm 0.9 \times10^{41}$ erg s$^{-1}$) using the relation in Kennicutt (1998): SFR = $7.9 \times 10^{-42}$ L(H$\alpha$) = $6.0\pm0.7 M_{\odot}$ year$^{-1}$. The halo accretion rate is therefore insufficient to produce the star formation rate of the galaxy. More relevant for galaxy formation, the halo accretion rate is two orders of magnitude too low to assemble the stellar mass of this galaxy within a Hubble time. If we preserve $\beta$ and $r_0$ for the halo, but increase $S_0$ to add the present-day stellar mass of $3.1\times10^{11} M_{\odot}$ to the halo, the cooling rate becomes $1.2-1.8 M_{\odot}$ year$^{-1}$, which is still insufficient to assemble the stellar mass by a factor of 20.

These results are also evidence against the emission around this galaxy being dominated by a galactic fountain or other internal processes related to star formation. One of the brightest known galactic fountains, in NGC 891, has a bolometric luminosity of $4\times10^{39}$ erg s$^{-1}$ (Bregman and Houck 1997) and a star formation rate of about $4 M_{\odot}$ year$^{-1}$ (Strickland et al. 2004b). For the NGC 1961 SFR of $6 M_{\odot}$ year$^{-1}$, the highest expected luminosity of a fountain is therefore $\sim 6\times10^{39}$ erg s$^{-1}$, whereas we measure an (absorbed) luminosity within 50 kpc of $2.6-3.0\times10^{40}$ erg s$^{-1}$. Moreover, the cooling time for the gas we observe beyond 18 kpc is greater than 10 Gyr, so even if the material has an internal origin, it should be treated as quasi-static instead of as a fountain. 

Within the context of galaxy formation models, the physical explanation for the missing baryons from NGC 1961 is unclear. The amount of missing matter in this galaxy and the depth of the potential well make it difficult for supernova-driven winds to expel the missing baryons. The escape velocity for material originating in the galactic disk (at an assumed radius of 10 kpc) in an NFW halo of mass $2.1\times10^{13} M_{\odot}$ is about $1300$ km s$^{-1}$. The missing baryonic mass is $3.2\times10^{12} M_{\odot}$, so the energy needed to unbind the missing baryons from the gravitational potential of the galaxy is $\sim 5\times10^{61}$ erg. A supernova with perfect coupling to the interstellar medium can inject up to $\sim 10^{51}$ erg of kinetic energy, so expelling the missing baryons from this galaxy requires $\sim 5\times10^{10}$ supernovae. Based on the present-day stellar mass of the galaxy and the average supernova rates per stellar mass per century from Mannucci et al. (2005), the expected total amount of supernovae (Type I + II) over 13 Gyr is $4\times10^9$, so NGC 1961 would need to have 13 times more stellar mass to supply enough energy from supernovae to unbind its missing baryons. Performing this calculation in more detail (e.g. accounting for smaller stellar mass at earlier times, allowing for cooling of supernova-heated gas) in general exacerbates the discrepancy. 

We also considered the possibility of AGN feedback for ejecting the missing baryons. However, NGC 1961 is a late-type spiral galaxy, so it has a relatively small bulge and presumably a correspondingly small central black hole, which makes an AGN-driven superwind less plausible. For example, if we assume the bulge contains a sixth of the stellar mass for this galaxy, and use the bulge mass-black hole mass relation (Marconi and Hunt 2003), we estimate a central black hole of mass $1\times10^8 M_{\odot}$. Assuming accretion converts $10\%$ of the infalling mass into energy, over its lifetime this black hole could have produced $2\times10^{61}$ erg, of which it is generally assumed only about 2\% can couple to the interstellar medium (Shankar et al. 2006, Bower et al. 2008) for a total available AGN feedback energy of only $4\times10^{59}$ erg. The most likely explanation for the missing baryons from this galaxy is probably some form of very early pre-heating which prevents the baryons from falling deeply into the potential well in the first place. 

We also note that other giant spirals show similar baryon deficits. For example, UGC 12591, at a distance of 134 Mpc (NED average) has a peak (inclination-corrected) HI rotation velocity of $476\pm23$ km s$^{-1}$ (Giovanelli \& Haynes 1985), which yields a virial mass of $3.5\times10^{13} M_{\odot}$ according to the NFW relation cited above. The HI mass is $4.3\times10^9 M_{\odot}$ (Ho 2007), and, using the 2MASS K$_S$-band total magnitude (8.9) as above we estimate a total stellar mass of $5.9\times10^{11} M_{\odot}$, for a baryon fraction of $0.017$ before including the hot halo. We also obtained XMM-Newton observations of the hot halo around this galaxy (Dai 2011, in preparation) which point to a similarly underluminous halo, to be discussed further in an upcoming paper.

In future work we hope to increase the sample of hot halos detected around giant spirals and further quantify the halo properties such as metallicity. If the trend of underluminous halos is common for the giant spirals, it would signal a break in the BTF somewhere between Milky-Way mass galaxies and the giant spirals.

\section{Acknowledgements}
The authors would like to thank the anonymous referee for a very thoughtful report which greatly improved the paper. We would also like to thank E. Bell, N. Calvet, X. Dai, A. Muratov, M. Ruszkowski, and C. Slater for helpful comments and discussions during the research and writing of this paper. We would also like to thank X. Dai for sharing with us his preliminary results and methods, S. Haan for helpfully sending his HI observations of this galaxy, and N. Bonaventura at the CXC help desk for assistance in reducing the Chandra data.  This research has made use of NASA's Astrophysics Data System. This research has made use of the NASA/IPAC Extragalactic Database (NED) which is operated by the Jet Propulsion Laboratory, California Institute of Technology, under contract with the National Aeronautics and Space Administration. This research has made use of software provided by the High Energy Astrophysics Science Archive Research Center (HEASARC), which is a service of the Astrophysics Science Division at NASA/GSFC and the High Energy Astrophysics Division of the Smithsonian Astrophysical Observatory. We gratefully acknowledge financial support from NASA through Chandra grant GO9-0089X. M.E.A. also acknowledges support from the NSF in the form of a Graduate Research Fellowship.

\section{References}
\noindent Anderson, M. E. and Bregman, J. N. 2010, ApJ 714:320\\\\
Bell, E. F. and de Jong, R. S. 2001, ApJ 550:212\\
Benson, A. J. et al. 2000, MNRAS 314:557\\
Benson, A. J. 2010, PhR 495:33\\
Bower, R. G. et al. 2006, MNRAS 370:645\\
Bower, R. G. et al. 2008, MNRAS 390:1399\\ 
Bregman, J. N. and Houck, J. C. 1997, ApJ 485:159\\
Combes, F. et al. 2009, A\&A 503:73\\
Combes, F., et al.\ 2009, \aap, 503, 73 \\
Crain, R. A. et al. 2010a, MNRAS 407:1403\\
Crain, R. A. et al. 2010b, MNRAS submitted, astro-ph/1011.1906\\
Dai, X. et al. 2011, in preparation\\
De Lucia, G. et al. 2006, MNRAS 366:499\\
Dekel, A. and Birnboim, Y. 2006, MNRAS 368:2\\
Dickey, J. M. and Lockman, F. J. 1990, ARA\&A 28:215\\
Dressler, A. 1980, ApJ 236:351\\
Dunkley, J. et al. 2009, ApJS 180:306\\
Forman, W., Jones, C., and Tucker, W. 1985, ApJ 293:102\\
Fukugita, M. and Peebles, P. J. E. 2006, ApJ 639:590\\
Giovanelli, R and Haynes, M. P. 1985, AJ 90:2445\\
Gottesman, S. T. et al. 2002, MNRAS 337:34\\ 
Grimm, H. J., Gilfanov, M., and Sunyaev, R. 2002, A\&A 391:923\\
Guedes, J. et al. 2011, submitted, astro-ph/1103.6030\\
Haan, S. et al. 2008, AJ 135:232\\
Ho, L. C. 2007, ApJ 668:94\\
Humphrey, P. J. et al. 2011, ApJ 729:53\\
Irwin, J. A., Athey, A. E., and Bregman, J. N. 2003, ApJ 587:356\\
Jarrett, T. H. et al. 2003, AJ 125:525\\
Kaufmann, T. et al. 2009, MNRAS 396:191\\
Kennicutt, R. 1998. ARA\&A 36:189\\
Kim, M. et al. 2007. ApJ 659:29\\
Klypin, A., Zhao, H. S., and Somerville, R. S. 2002, ApJ 573:597\\
Li, Z. et al. 2006, MNRAS 371:147\\
Li, Z., Wang, Q. D., and Hameed, S. 2007, MNRAS 376:960\\
O'Sullivan, E., Forbes, D. A., and Ponman, T. J. 2001, MNRAS 328:461\\
Owen, R. A. and Warwick, R. S. 2009, MNRAS 394:1741\\
Mannucci, F. et al. 2005, A\&A 433:807\\
Marconi, A. and Hunt, L. K. 2003, ApJ 589:21L\\
McGaugh, S. S. et al. 2010, ApJ 708:14\\
Mulchaey, J. S., and Jeltema, T. E. 2010, ApJ 715:L1\\
Navarro, J. F. 1998, preprint (astro-ph/9807084)\\
Pedersen, K. et al. 2006, New Astron. 11:465\\
Rasmussen, J. et al. 2009, ApJ 697:79\\
Read, A. M. and Ponman, T. J. 1998, MNRAS 297:143\\
Revnivtsev, M. et al. 2008. A\&A 490:37\\
Roberts, T. P. and Warwick, R. S. 2000, MNRAS 315:98\\
Robertson, B. et al. 2006, ApJ 645:896\\
Rubin, V. C. et al. 1979, ApJ 230:35\\
Sarazin, C. L. 1986, RevModPhys 58:1\\
Shankar, F. et al. 2006, ApJ 643:14\\
Shattow, G. and Loeb, A. 2009, MNRAS 392:L21\\
Sommer-Larsen, J. 2006, ApJ 644:L1\\
Stark, D. V., McGaugh, S. S., and Swaters, R. A. 2009, AJ 138:392\\
Strickland, D. K. et al. 2004a, ApJS 151:193\\
Strickland, D. K. et al. 2004b, ApJ 606:829\\
Sutherland, R. S. and Dopita, M. A. 1993, ApJS 88:253\\
T\"{u}llmann, R. et al. 2006, A\&A 448:43\\
Voss, R. et al. 2009, ApJ 701:471\\
White, S. D. M. and Frenk, C. S. 1991, ApJ 379:52\\
White, S. D. M. and Rees, M. J. 1978, MNRAS 183:341\\
Yamasaki, N. Y. et al. 2009, PASJ  61:291\\
Guedes, J., Callegari, S., Madau, P., \& Mayer, L.\ 2011, arXiv:1103.6030

%\section{Figures}

\begin{figure}
\plotone{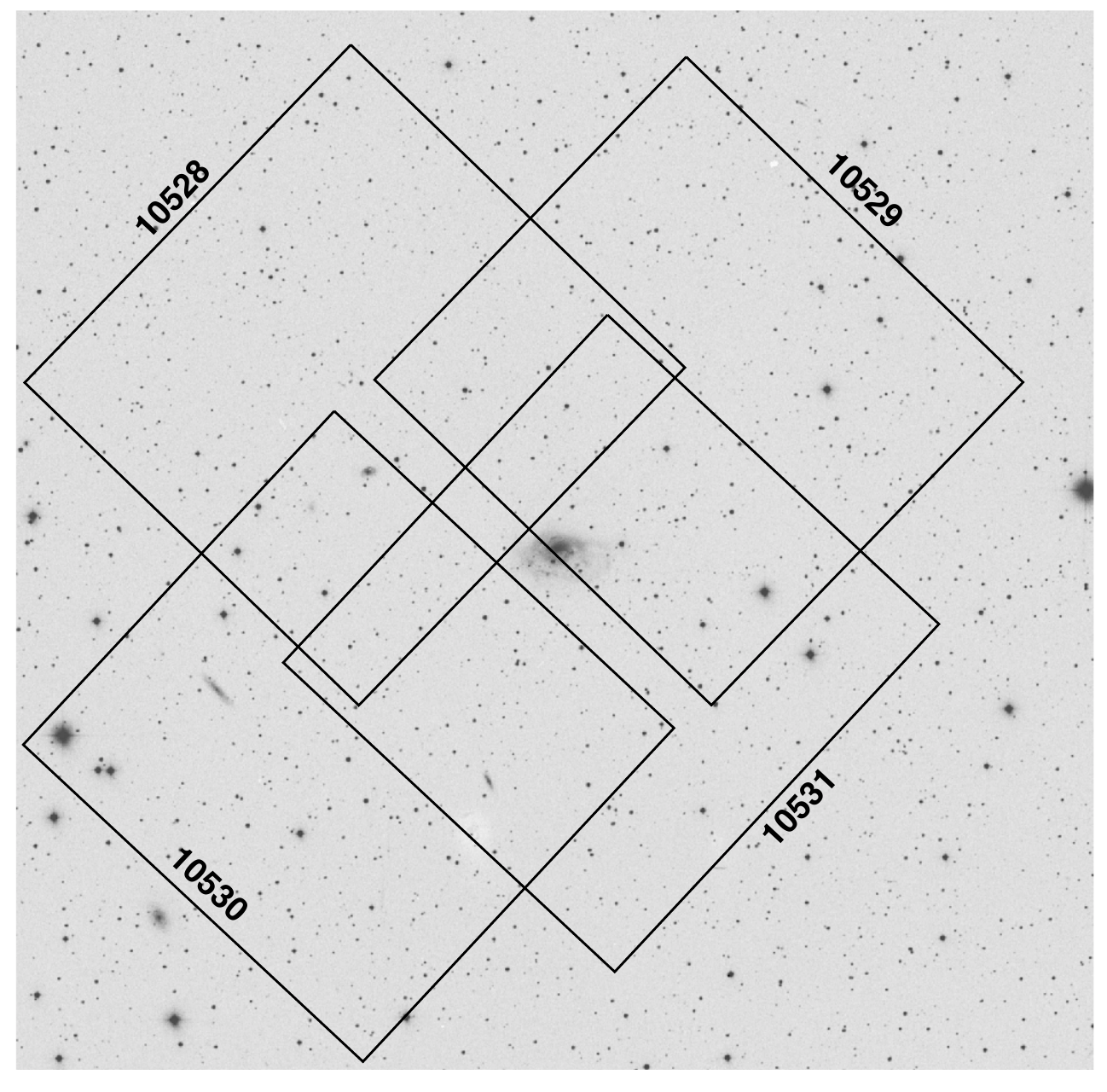}
\caption{\small An ESO DSS image of NGC 1961, with the layout of our four ACIS-I observations overlaid. North is up, East is left, and each box is about 17 arcmin (280 kpc) on a side. }
\end{figure}

\begin{figure}
\plottwo{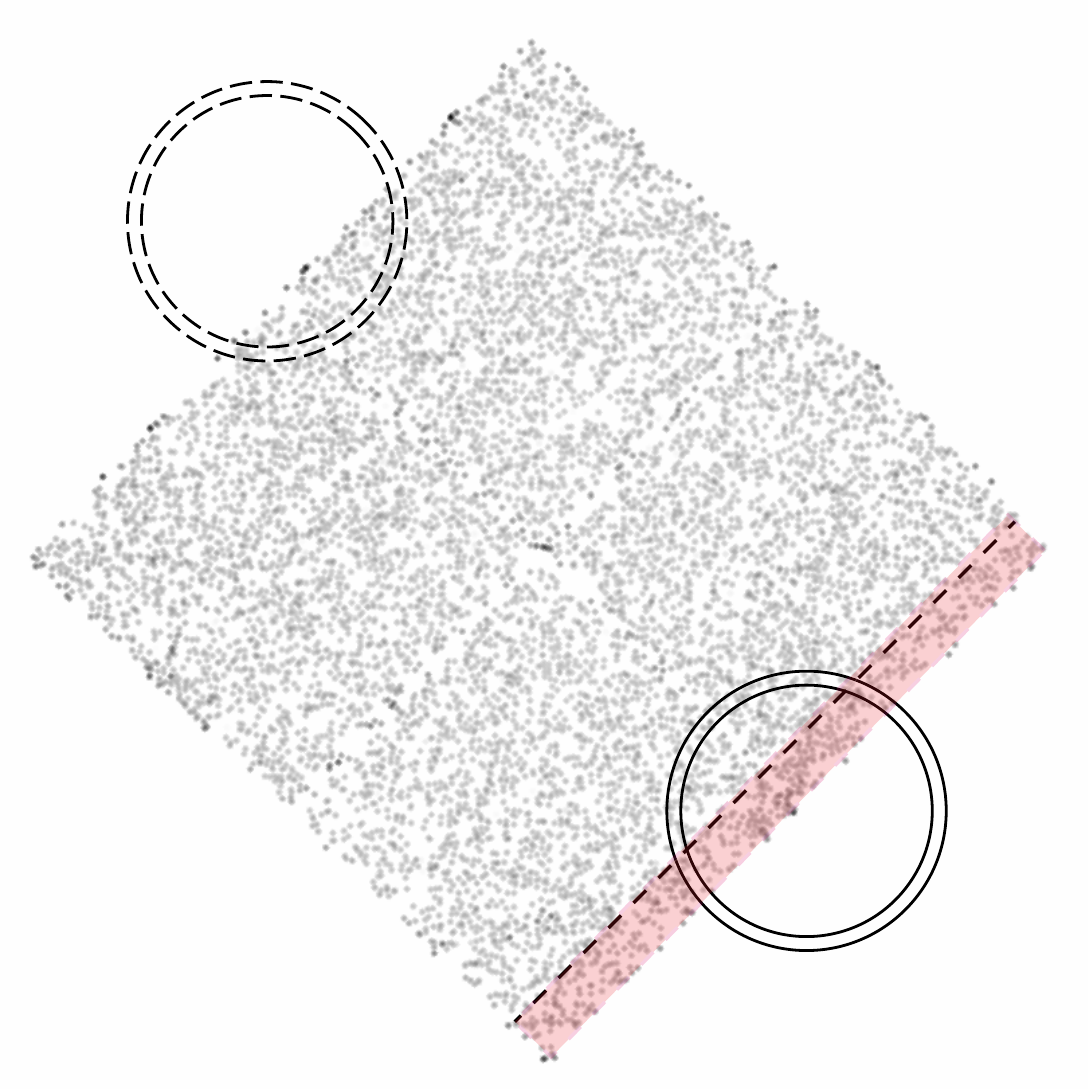}{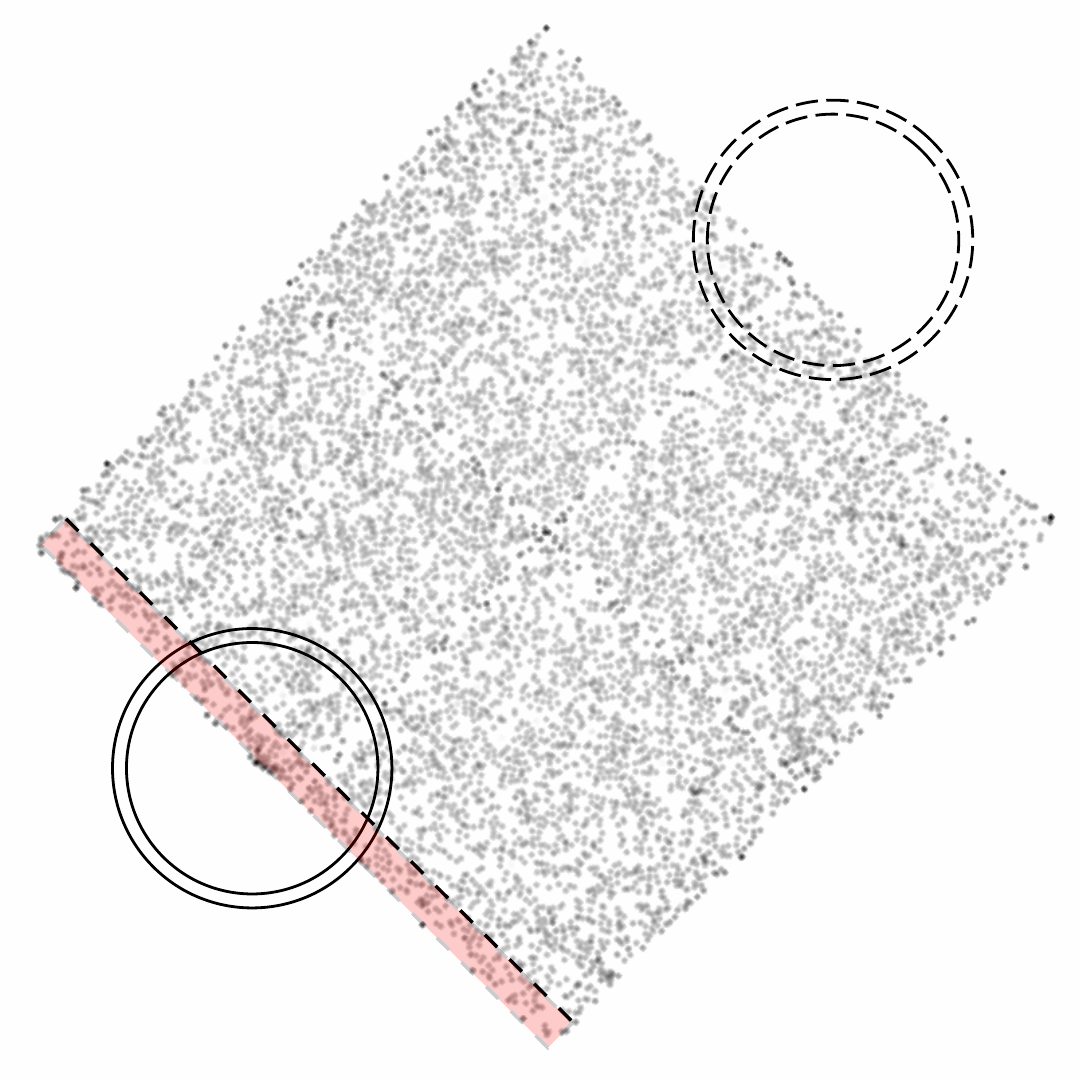}
\plottwo{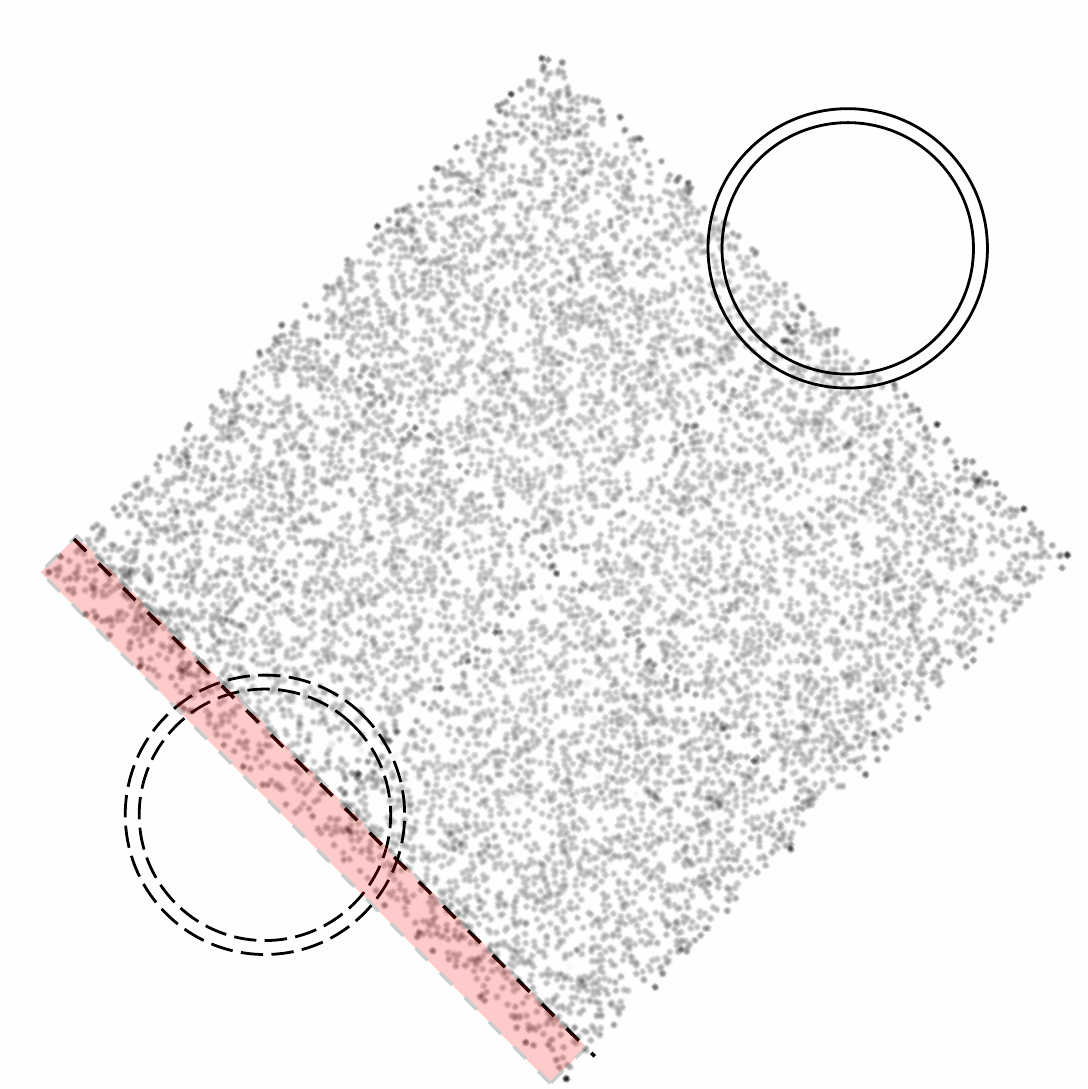}{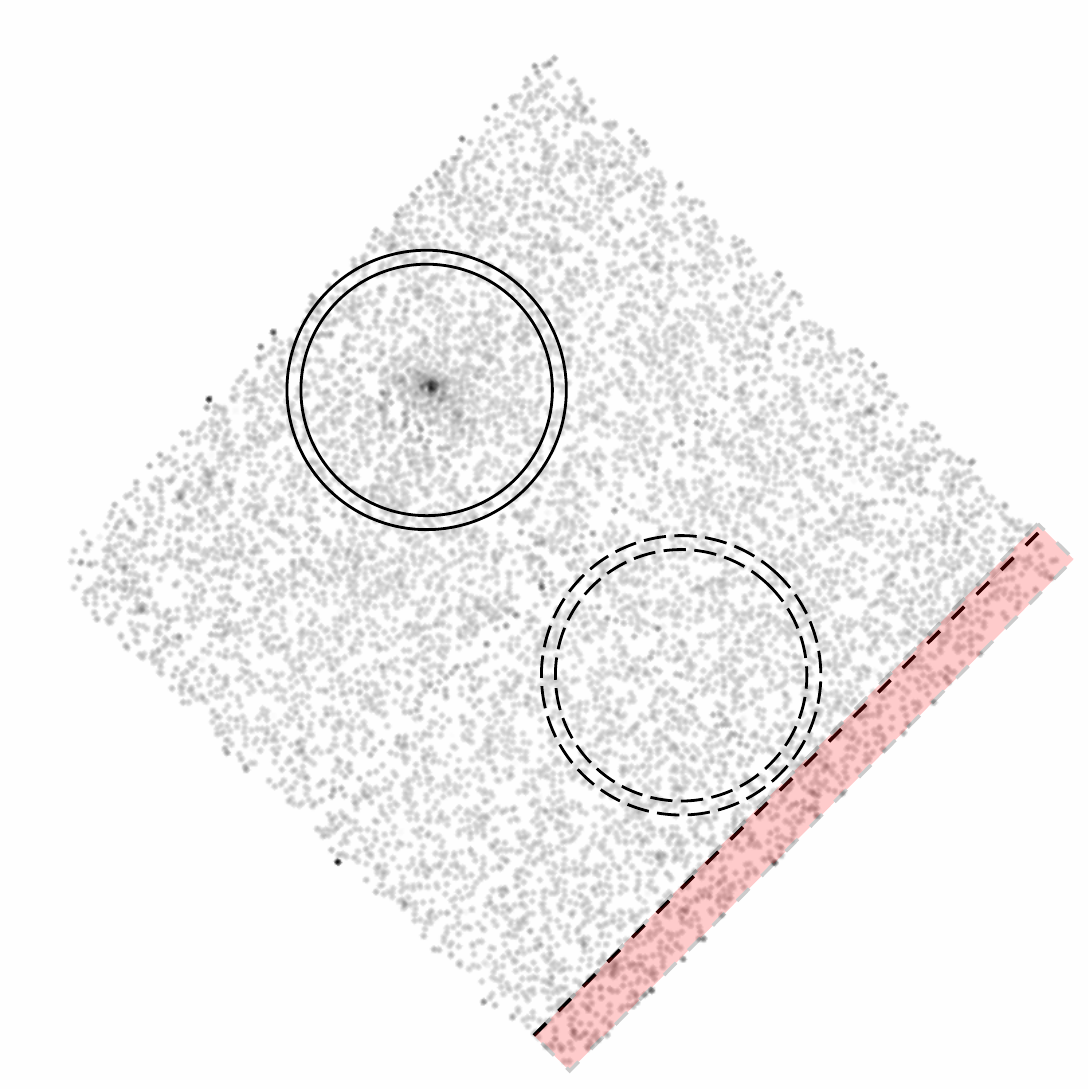}
\caption{\small Our four Chandra ACIS-I observations of NGC 1961. Observation 10528 is top left, observation 10529 is top right, observation 10530 is bottom left, observation 10531 is bottom right. Each image uses 5-pixel Gaussian smoothing and square root scaling, and North is up and East is left. The annuli show our in-field conjugate background subtraction technique for measuring a background-subtracted surface brightness profile around NGC1961. We use annuli around the source and subtract background from the annulus of equivalent size on the opposite side of the aimpoint. The solid annulus denotes an example source region, and the dashed annulus an example conjugate region. The red shaded region represents the area on each detector for which we were unable to subtract a conjugate background from the $180^{\circ}$ annulus; for data within these regions in observations 10528 and 10529 we used conjugate regions at $120^{\circ}$ and $240^{\circ}$ instead. For observation 10531, at radii larger than 4.5 arcseconds the annuli cross each other, so we cannot use points at larger radius.  }
\end{figure}

\begin{figure}
\plotone{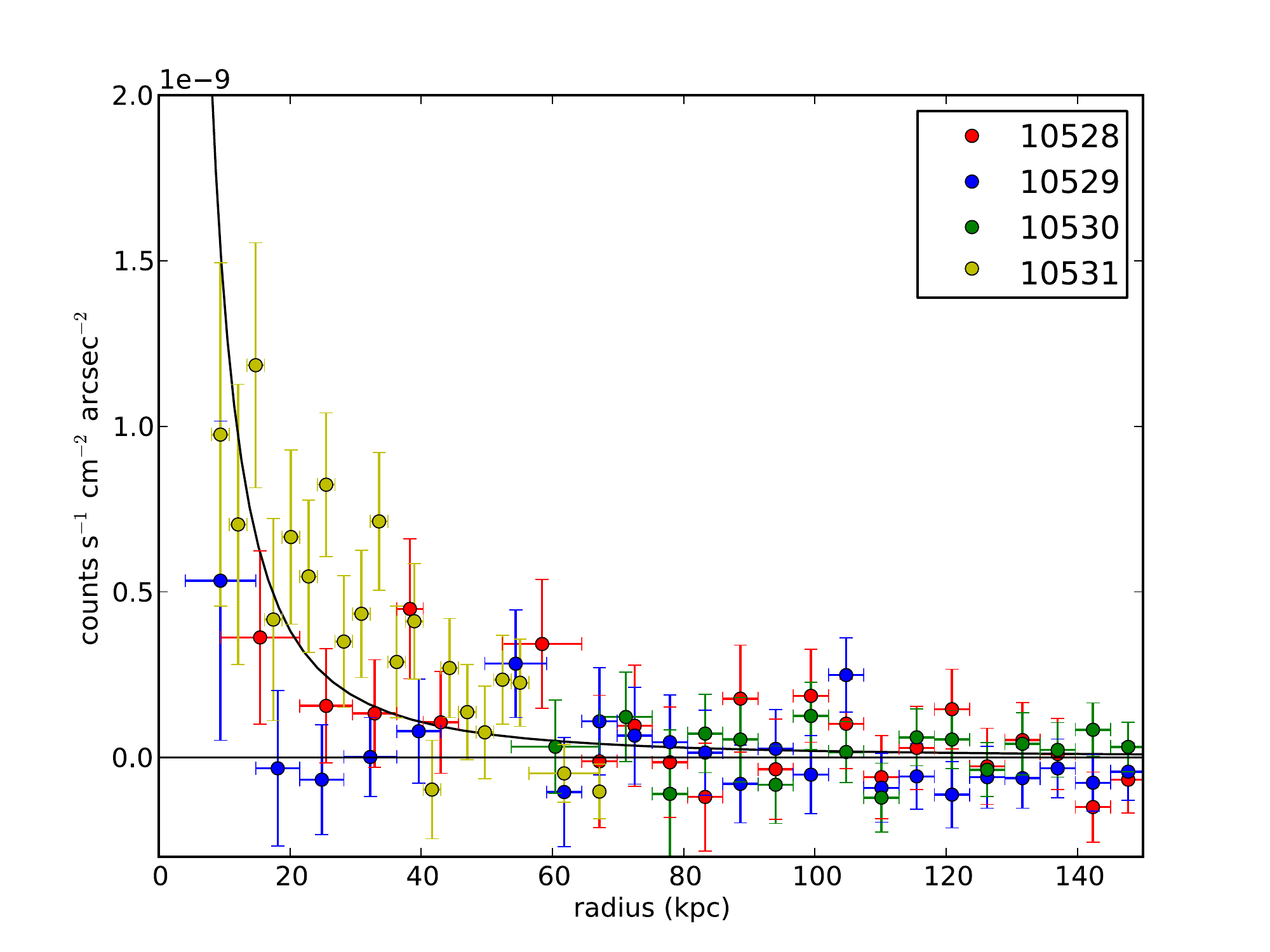}
\caption{\small Background-subtracted radial surface brightness profile of NGC 1961, including binned data from all four observations in the same figure. The black line is the best-fit $\beta$-model to the data. The model has parameters $S_0 = 9.77\times10^{-8}$ count s$^{-1}$ cm$^{-2}$ arcsec$^{-2}$, $r_0 = 1.00$ kpc, $\beta = 0.47$. The $\chi^2$ is 66.9 for 50 degrees of freedom, so the fit is inconsistent with the data at less than 95\% confidence. }
\end{figure}

\begin{figure}
\plotone{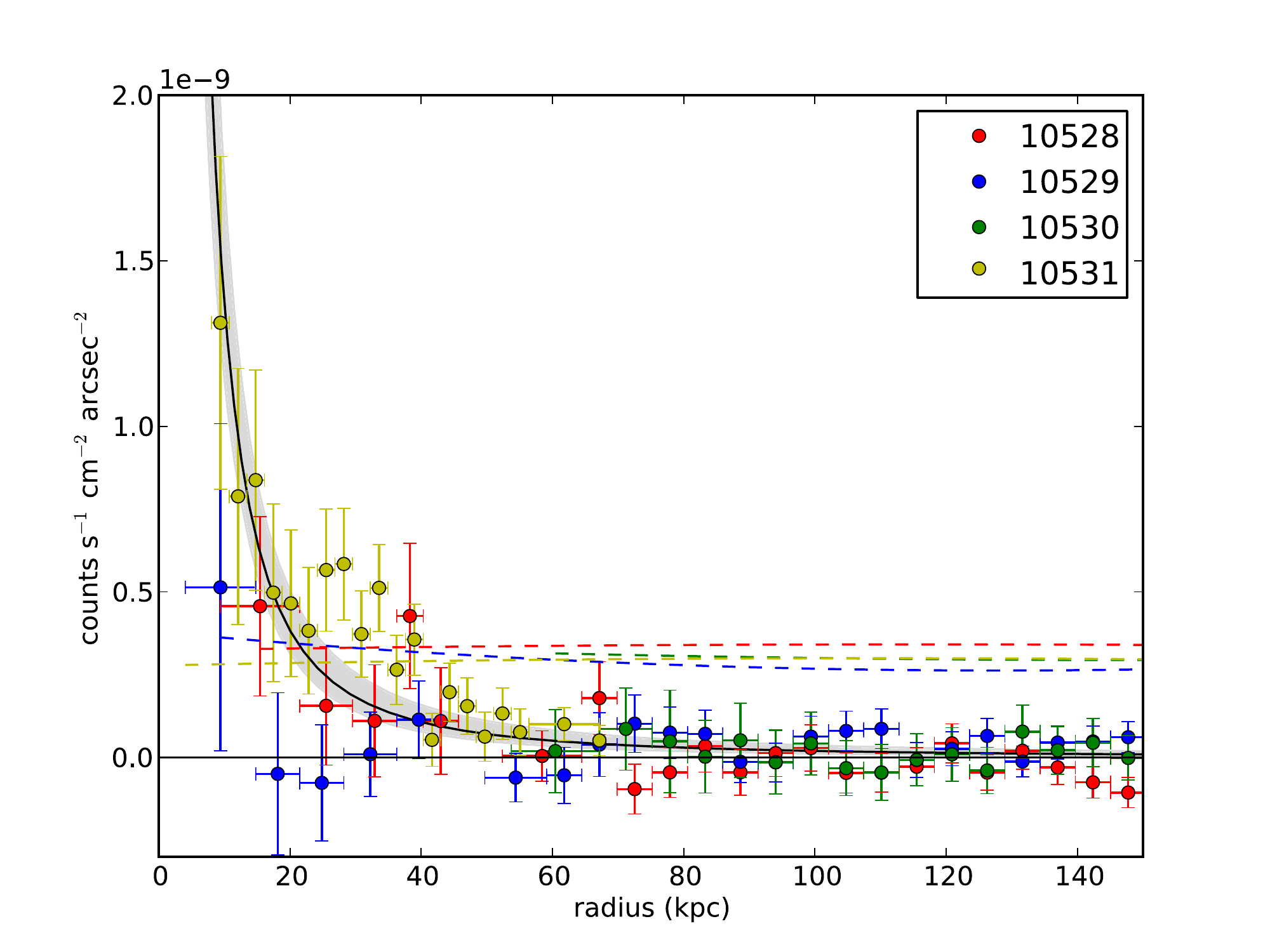}
\caption{\small Background-subtracted radial surface brightness profiles for all four observations after background smoothing (see text). The smoothed fits to the background are represented by the colored dashed lines (the dotted lines begin where the background data are extrapolated beyond the edge of the detector). The black line is the best-fit $\beta$-model to the data. The model has parameters $S_0 = 8.90\times10^{-8}$ count s$^{-1}$ cm$^{-2}$ arcsec$^{-2}$, $r_0 = 1.00$ kpc, $\beta = 0.46$. The $\chi^2$ is 62.1 for 50 degrees of freedom, so the fit is inconsistent with the data at less than 95\% confidence. The gray shaded region denotes all the acceptable fits to the data at this confidence level; we use this region as the fiducial range throughout the paper. }
\end{figure}

\begin{figure}
\plotone{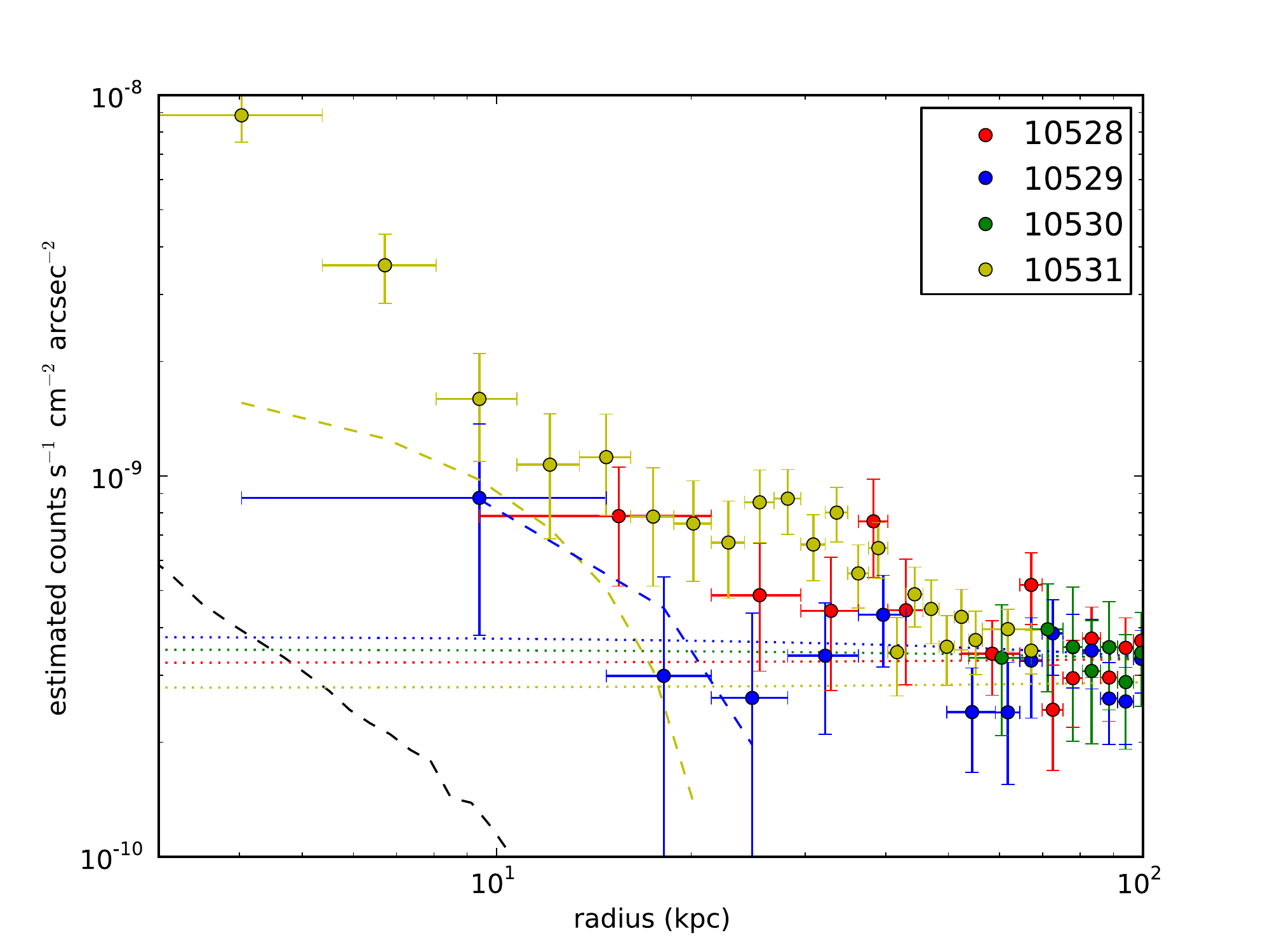}
\caption{\small Log-log plot of radial surface brightness profiles for all four observations. The black dashed line is the estimated contribution of stars, and the colored dashed lines are the estimated contributions of X-ray binaries. The colored data points are the surface brightness profile with resolved and unresolved point sources subtracted. Unlike Figures 3 and 4, we have not subtracted the sky X-ray background from the surface brightness profile. The smoothed sky X-ray background is indicated by the four dotted colored lines. We detect emission above the background out to 40-50 kpc which is more spatially extended than the other galactic components. }
\end{figure}

%\begin{figure}
%\plotone{figures/flat_chi2.pdf}
%\caption{\small Results of fitting a two-component $\beta$-model to each observation. One component is a "flattened" profile with $\beta = 0.35$ and $r_0 = 50$ kpc; the choices of normalization are displayed on the X-axis. The other component has all three parameters free; the $\chi^2$ of the fit to the choice of parameters that minimize $\chi^2$ is shown on the Y-axis. If there was no fit inconsistent to the data at less than 95\%, the $\chi^2$ is set to the value corresponding to a 95\% confidence level. We find an upper limit on $S_0$ of $1.5\times10^{-11}$ count s$^{-1}$ cm$^{-2}$ arcsec$^{-2}$. }
%\end{figure}

\end{document}